\documentclass{emulateapj}
\usepackage{hyperref} 
\usepackage{natbib}
\usepackage{graphicx}

\newcommand{\myemail}{DJcosmic@gmail.com}

\begin{document}

\title{Cyclostationary in The Time Variable Universe}  

\author{Jiang Dong\altaffilmark{1}}
\affil{Yun Nan Astronomical Observatory, NAOs, CAS}
\email{\myemail}

\begin{abstract}
Cyclostationary processes are those signals
whose have vary almost periodically in statistics.
It can give rise to random data 
whose statistical characteristics vary periodically with time 
although these processes not periodic functions of time.
Intermittent pulsar is a special type in pulsar astronomy 
which have period but not a continuum.
The Rotating RAdio TransientS (RRATs) represent 
a previously unknown population of bursting neutron stars.
Cyclical period changes of variables star also can be thought as cyclostationary
which are several classes of close binary systems.
Quasi-Periodic Oscillations (QPOs) refer to the way the X-ray light 
from an astronomical object flickers about certain frequencies 
in high-energy (X-ray) astronomy.
I think that all above phenomenon is cyclostationary process. 
I describe  the signal processing of cyclostationary,
then  discussed that the relation 
between it and intermittent pulsar, RRATs, cyclical
period changes of variables star and QPOs,
and give the perspective of finding the cyclostationary source 
in the transient universe.
\end{abstract}


\section{Introduction}
It was first mentioned by the design of synchronization algorithms 
for communications systems that cyclostationary process\citep{spsg05}.
Many processes encountered in nature arise from periodic phenomena.
For example, in telecommunications, telemetry, radar, and sonar applications, 
periodicity is due to modulation, sampling, multiplexing, and coding operations.
In mechanics it is due, for example, to gear rotation.
In econometrics, it is due to seasonality; 
and in atmospheric science it is due to rotation and revolution of the earth
\citep{gnp06,spsg05,g94}.
Cyclostationary processes are named in multiple different ways 
such as periodically correlated, periodically nonstationary, 
periodically nonstationary or cyclic correlated processes 
in literature\citep{h89,h97}.

In astronomy, now just in removing cyclostationary radio frequency interferences
(RFI) from radio astronomical data\citep{lvb00,bw05}
and the galactic white-dwarfs background that will be observed 
by LISA\citep{e05} cyclostationary process be mentioned.
Some radio pulsar have been discovered the phenomenon of 
intermittent period in recently\citep{klo+06}.
The nulling of the pulsar be thought a characteristic of the older pulsars.
In X-ray pulsar, the phenomenon of intermittent
far-flung exist for the effect of companion.
It also show almost period in time series.
The Rotating RAdio TransientS (RRATs) demonstrate
a previously unknown population of bursting neutron stars\citep{mll+06}.
Cyclical period changes of variables star 
that including Algol, WUrsaeMajoris, and RS Canum Venaticorum systems 
and the cataclysmic variables, and RR Lyrae star et al. 
also can be thought as cyclostationary.
The fastest variability components in X-ray binaries are 
the kilohertz quasi-periodic oscillations (kHz QPOs), which occur in 
a wide variety of low magnetic-field neutron star systems\citep{vdk05}.
The signal process of the above phenomenon,
in a statistical sense, as a periodic function of time. 
These kind of processes have been studied for many years, 
and are usually referred to as cyclostationary random processes
(see\citep{gnp06,spsg05,g94}
for a comprehensive overview of the subject and for more references). 

In what follows we will briefly summarize the properties 
of cyclostationary processes,
then discuss the cyclostationary signal process in some astronomical phenomenon,
in the end give the perspective of finding cyclostationary processes   
in the time variable universe.

\section{The basic of  cyclostationary processes}
The continuous stochastic process ${\cal X} (t)$ having 
finite second order moments is said to be {\em cyclostationary\/} with 
period $T$ if the following expectation values
\begin{eqnarray}
E[{\cal X} (t)] &=& m(t) = m(t + T) , \\
E[{\cal X} (t') {\cal X} (t)] &=& C(t',t) = C(t' + T,t + T)
\end{eqnarray}
are periodic functions of period $T$, 
for every $(t',t) \in {\bf R} \times {\bf R}$. 
We will assume $m(t) = 0$ for simplicity now\citep{gf75,h89,h97}.
The important special case of cyclostationary signals are 
those which exhibit cyclostationary in second-order statistics 
(e.g., the autocorrelation function). 
It is called wide-sense cyclostationary signals, 
and is analogous to wide-sense stationary processes. 
The exact definition differs depending on whether the signal is treated as
a stochastic process or as a deterministic time series\citep{g78}.

If ${\cal X} (t)$ is cyclostationary, 
then the function $B(t,\tau) \equiv C(t + \tau,t)$ for 
a given $\tau \in {\bf R}$ is periodic with period $T$,
and it can be represented by the following Fourier series
\begin{equation}
B(t,\tau) = \sum_{r = -\infty}^{\infty}B_r(\tau) e^{i2\pi\frac{r t}{T}} \ ,
\end{equation}
where the functions $B_r(\tau)$ are given by
\begin{equation}
B_r(\tau) = \frac{1}{T}\int^T_0B(t,\tau) e^{- i 2\pi r\frac{t}{T}} \ dt \ .
\label{eq:FB}
\end{equation}
The Fourier transforms $g_r(f)$ of $B_r(\tau)$ are the so called
``cyclic spectra'' of the cyclostationary process ${\cal X} (t)$\citep{h89}
\begin{equation}
g_r(f) = \int_{-\infty}^{\infty}B_r(\tau)e^{-i 2\pi f \tau } \ d\tau \ .
\end{equation}
If a cyclostationary process is real, 
the following relationships between the cyclic spectra hold
\begin{eqnarray}
B_{-r}(\tau) & = & B^*_r(\tau) \ ,
\label{eq:ksym1}
\\
g_{-r}(-f) & = &  g^*_r(f) \ ,
\label{eq:ksym2}
\end{eqnarray}
where the symbol $^*$ means complex conjugation.  
This implies that, for a real cyclostationary process, 
the cyclic spectra with $r \geq 0$
contain all the information needed to characterize the process itself.

The function $\sigma^2(\tau) \equiv B(0,\tau)$ is 
the variance of the cyclostationary process ${\cal X} (t)$, 
and it can be written as a Fourier decomposition as 
a consequence of Eq. (\ref{eq:FB})
\begin{equation}
\sigma^2(\tau) = \sum_{r=-\infty}^{\infty}H_r e^{ i
2\pi\frac{r \tau}{T}},
\end{equation}
where $H_r \equiv B_r(0)$ are harmonics of the variance $\sigma^2$.
From Eq. (\ref{eq:ksym1}) it follows that $H_{-r} = {H}^*_r$.

For a discrete, finite, real time series ${\cal X}_t$, $t = 1,\ldots, {\cal N}$ 
the cyclic spectra can be estimated by 
generalizing standard methods of spectrum estimation used with
stationary processes. 
Assuming again the mean value of the time series ${\cal X}_t$ to be zero, 
the cyclic autocorrelation sequences are defined as
\begin{equation}
s_l^r = \frac{1}{\cal N}\sum_{t=1}^{{\cal N}-|l|}{\cal X}_t {\cal X}_{t+|l|}
e^{-\frac{i 2\pi r (t-1)}{T}} \ . 
\label{eq:cycorr}
\end{equation}
The cyclic autocorrelations are
asymptotically (i.e.\ for $N \rightarrow \infty$) unbiased estimators
of the functions $B_r(\tau)$\citep{h89}. 
The Fourier transforms of the cyclic autocorrelation sequences $s_l^r$ 
are estimators of the cyclic spectra $g_r(f)$. 
These estimators are asymptotically unbiased, 
and are called ``inconsistent estimators'' of the cyclic spectra, 
i.e. their variances do not tend to zero asymptotically.  
In the case of Gaussian processes\citep{h89} consistent estimators 
can be obtained by first applying a lag window to 
the cyclic autocorrelation and then perform a Fourier transform.  

The alternative procedure for identifying consistent estimators of 
the cyclic spectra is to first take the Fourier transform\citep{gnp06,g94},
$\tilde{{\cal X}}(f)$, of the time series ${\cal X} (t)$
\begin{equation}
\tilde{{\cal X}}(f) = \sum_{t = 1}^{\cal N}  {\cal X}_t e^{-i 2\pi f (t - 1)}
\end{equation}
and then estimate the cyclic periodograms $g_r(f)$
\begin{equation}
g_r(f) = \frac{\tilde{{\cal X}}(f)\tilde{{\cal X}}^*(f - \frac{2\pi
    r}{T})}{\cal N} \ .
\end{equation}
By finally smoothing the cyclic periodograms, 
consistent estimators of the spectra $g_r(f)$ are then obtained.  
The estimators of the harmonics $H_r$ of 
the variance $\sigma^2$ of a cyclostationary random process can 
be obtained by first forming a sample variance of the time series ${\cal X}_t$. 
The sample variance is obtained by dividing the time series ${\cal X}_t$ into 
contiguous segments of length $\tau_0$ such that $\tau_0$ is 
much smaller than the period $T$ of the cyclostationary process, 
and by calculating the variance $\sigma^2_I$ over each segment.  
Estimators of the harmonics are obtained either by Fourier analyzing 
the series $\sigma^2_I$ or by making a least square fit to $\sigma^2_I$ 
with the appropriate number of harmonics. 
Note that the definitions of 
(i) zero order ($r = 0$) cyclic autocorrelation, 
(ii) periodogram, and 
(iii) zero order harmonic of the variance, 
coincide with those usually adopted for stationary random processes. 
Thus, even though a cyclostationary time series is not stationary, 
the ordinary spectral analysis can be used for obtaining the zero order spectra.
Note, however, that cyclostationary random processes provide 
more spectral information about the time series they are associated with 
due to the existence of cyclic spectra with $r > 0$\citep{e05}.

For stationary and cyclostationary time-series, 
there is two alternative philosophical frameworks 
for the two problems of estimating the time-invariant 
or time-variant autocorrelation function and its Fourier transform.
One is based on the stochastic process model, 
and the other is based on the nonstochastic time-series model. 
Gardner, W.A. compared it, 
and explained that results on estimator bias and variance 
for these two problems couched within the stochastic process framework have
analogs within the nonstochastic framework. 
The bias and variance results for cyclostationary time-series that are
available within these two frameworks\citep{g91}.

As an important and practical application, 
assuming consider a time series $y_t$ consisting of 
the sum of a stationary random process, $n_t$, 
and a cyclostationary one ${\cal X}_t$ (i.e. $y_t = n_t + {\cal X}_t$).  
Let the variance of the stationary time series $n_t$ be $\nu^2$ 
and its spectral density be $\mathcal{E}(f)$.  
It is easy to see that the resulting process is also cyclostationary. 
If the two processes are uncorrelated, 
then the zero order harmonic $\Sigma^2_0$ of the variance of 
the combined processes is equal to
\begin{equation}
\Sigma^2_0 = \nu^2 + \sigma^2_0 \ ,
\end{equation}
and the zero order spectrum, $G_0(f)$, of $y_t$ is
\begin{equation}
G_0(f) =  \mathcal{E}(f)  +  g_0(f) \ .
\end{equation}
The harmonics of the variance as well as the cyclic spectra of $y_t$
with $r > 0$ coincide instead with those of ${\cal X}_t$.  
In other words, the harmonics of the variance 
and the cyclic spectra of the process $y_t$ with $r > 0$ contain information 
only about the cyclostationary process ${\cal X}_t$, 
and are not ``contaminated'' by the stationary process$n_t$\citep{e05}. 

\section{ Cyclostationary in Pulsar, QPOs
and Variables Star }
\begin{figure}
\epsscale{.80}
\plotone{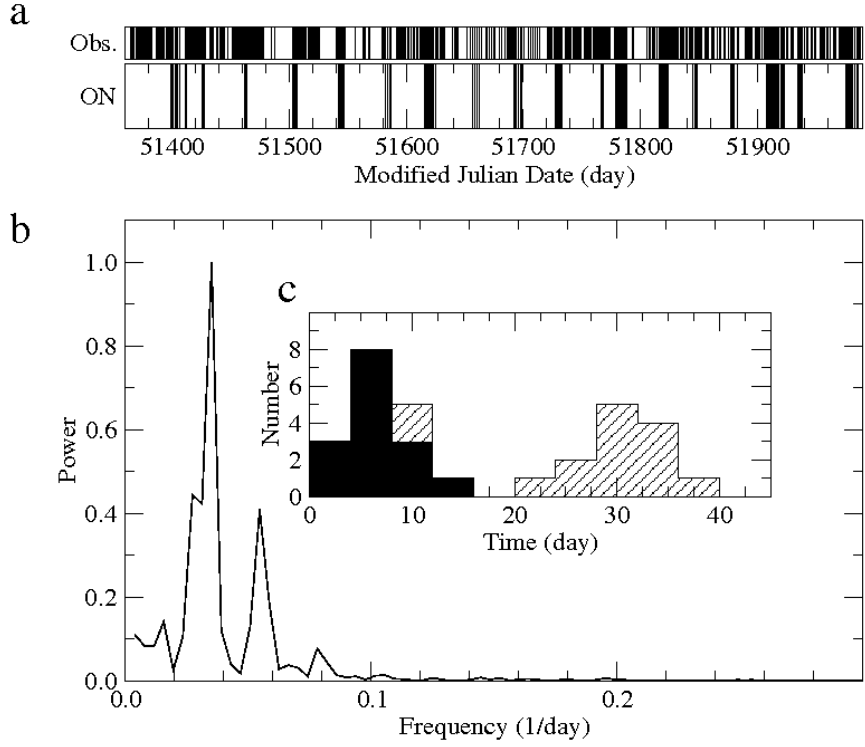}
\caption{Time variation of the radio emission of PSR~B1931+24.
During the on phases, the pulsar is easy to detect and has the
stable long-term intrinsic flux density associated with most normal
pulsars. Since 1998, the pulsar has been observed as frequently as
twice a day.\citep[see][]{klo+06}}
 \label{fig:IP}
\end{figure}

Recently, Kramer, M. et al. discovered one class of neutron stars 
that are seemingly ordinary radio pulsars, 
but which are only active for some short time and in a quasi-periodic fashion. 
So they call these "Intermittent Pulsars". 
These pulsar that is only periodically active. 
It appears as a normal pulsar for about a week 
and then "switches off" for about one month
before emitting pulses again. The pulsar, called PSR B1931+24, 
is unique in this behaviour and affords astronomers an opportunity to
compare its quiet and active phases.
Most surprisingly, the pulsar rotation slows down faster 
when the pulsar is on than when it is off\citep{klo+06}.
In the Figure \ref{fig:IP}, 
part a) show a typical sequence of observations covering a 20-month
interval is indicated by the black lines. 
It shows respectively the times of observation and the times when
PSR~B1931+24 was on.
It is clear that the pulsar is not visible for $\sim$80\% of the time.  
part b) give the appearance of the pulsar is quasi-periodic nature, 
demonstrated by the power spectrum of the intensity obtained 
from the Fourier Transform of the autocorrelation function of 
the mean pulse flux density obtained over the same 20-month interval. 
part c) is histograms of 
the durations of the on (solid) and off (hatched) phases.  
In off phases, integration over several weeks shows that 
any pulsed signal has 
a mean flux density of less than 2~$\mu$Jy at 1400~MHz\citep{klo+06}.. 

\begin{figure}
\epsscale{.60}
\plotone{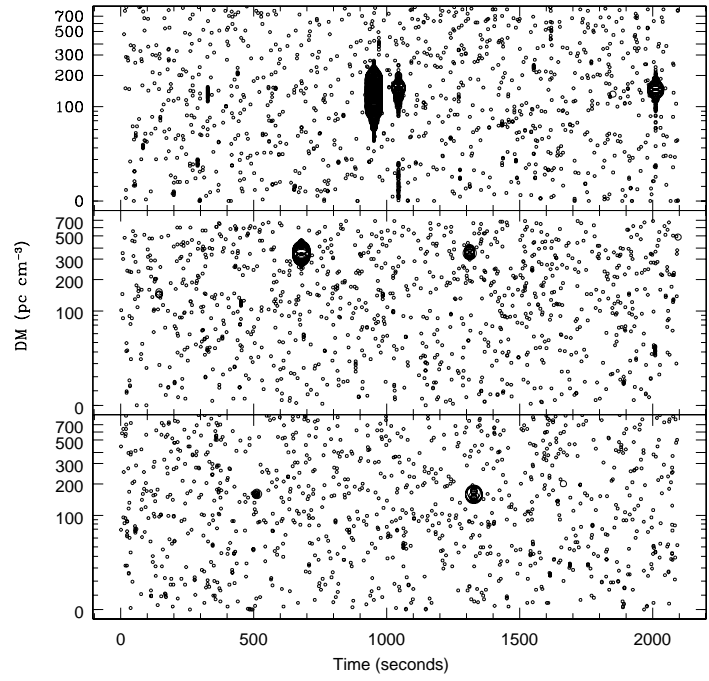}
\caption{{\bf The observational signatures of the new radio
transient sources.} 
\citep[see][]{mll+06}}
 \label{fig:RRATs}
\end{figure}

McLaughlin, M.A. et al. discovered a class of Rotating Radio Transients(RRATS) 
which are identified as rotating neutron stars that 
send out very short flashes of radio light. 
These flashes are very short and very rare: one hundredth of a second long,
the total time the objects are visible amounts to 
only about one tenth of a second per day. 
The isolated flashes last for between 2 and 30 milliseconds. 
In between, for times ranging from 4 minutes to 3 hours,
the new stars are silent.
They current estimates suggest that these objects are
four times more common in the Galaxy than radio pulsars.
In the Figure \ref{fig:RRATs}, from top to bottom, 
it is show that the original detections of J1317--5759, J1443--60 
and J1826--14 in the Parkes Multibeam Survey data\citep{mll+06}.
It is show that RRATs still have period in short time scale.
They have identified periodicities 
in the range of 0.4 to 7 s in 10 of the 11 objects\citep{lyne07}.

Camilo, F. et al. show that XTE J18102197 emits bright, narrow, highly
linearly polarized radio pulses, observed at every rotation, thereby
establishing that magnetars can be radio pulsars. 
There is no evidence of radio emission before the 2003 X-ray outburst 
(unlike ordinary pulsars, which emit radio pulses all the time), 
and the flux varies from day to day. 
The flux at all radio frequencies is approximately equal-and at $>$ 
20GHz\citep{crh+06}.

In wide-sense, if one pulsar have nulling or giant pulse phenomenon,
it will not have strict period function with time,
just in statistics have periodic.
Another pulsar cyclostationary process is Shabanova, T.V. find that 
the nature of the observed cyclical changes in the timing residuals 
from PSR B1642 – 03 is a continuous generation of peculiar glitches 
whose amplitudes are modulated by a periodic large-scale sawtooth-like
function.  As the modulation function is periodical, 
the picture of cyclical timing residuals will be exactly
repeated in each modulation period or every 60 years\citep{s09}.

\begin{figure}
\includegraphics[angle=270,scale=.30]{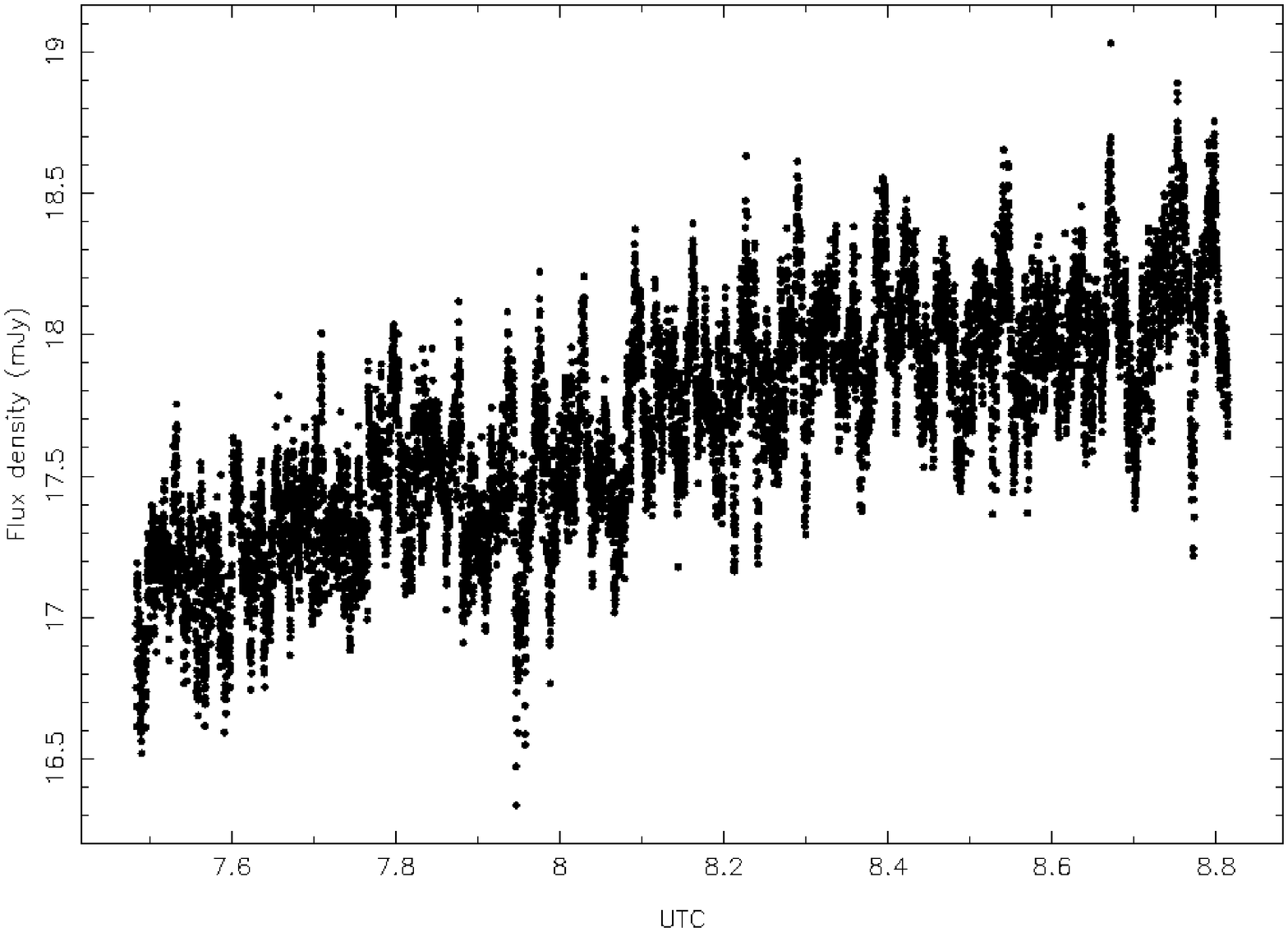}
\vspace{5mm}
\includegraphics[angle=270,scale=.30]{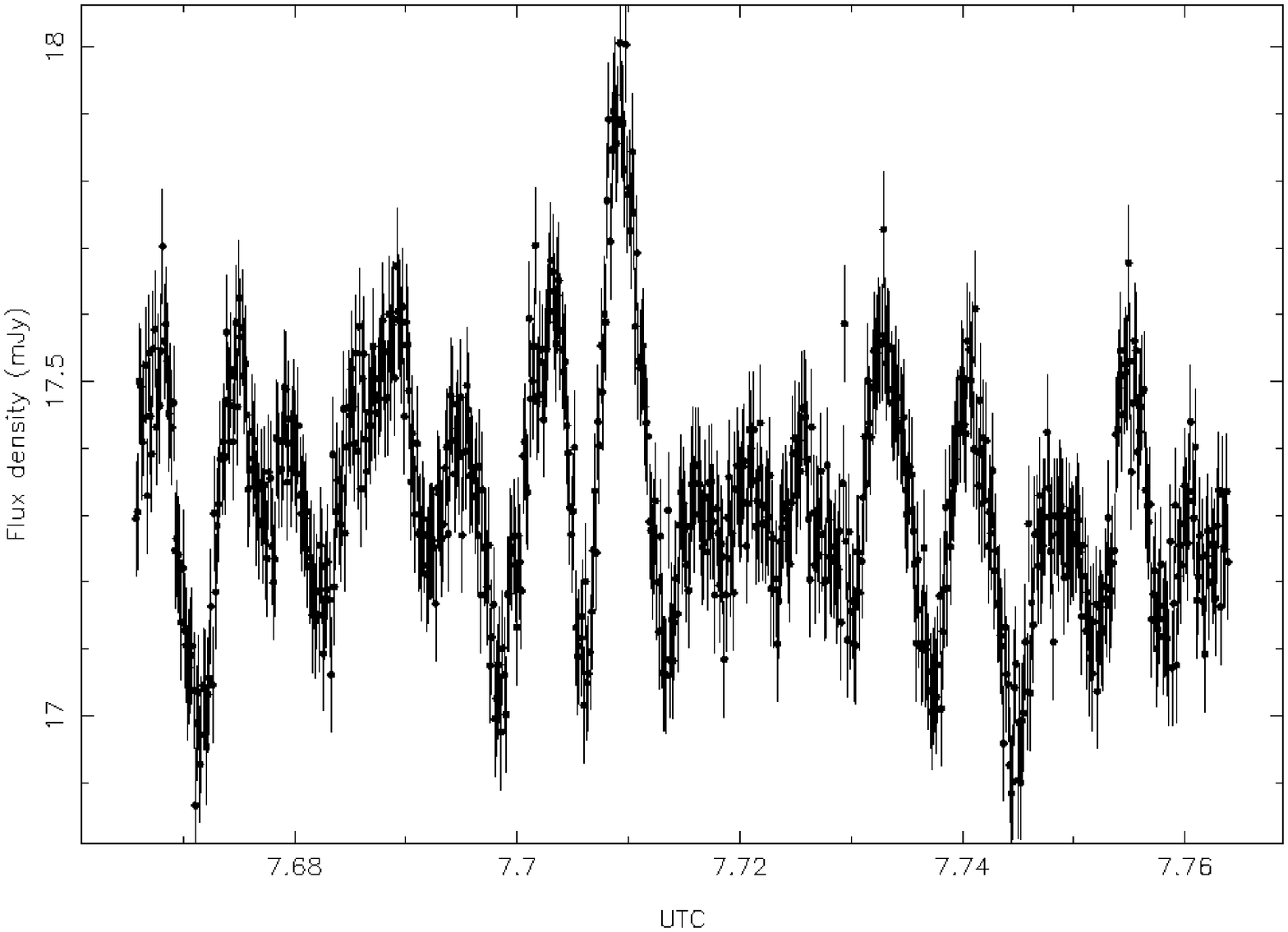}
\caption{Left panel, lightcurve of the optical variability in Hercules
  X-1 for the entire data-set. 
Right panel, lightcurve of a subset of the data clearly showing the QPO. 
\citep{ohb+01}}
\label{fig:QPO}
\end{figure}

QPOs were first identified in white dwarf systems\citep{QPO85} 
and then in neutron star systems\citep{vdk05}.
Two QPO peaks (the 'twin peaks') occur in 
the power spectrum of the X-ray flux variations. 
They move up and down in frequency together in the 300-1200Hz range in 
correlation with source state and often, luminosity. 
The typically 300-Hz peak separation usually decreases by a few tens
of Hz when both peaks move up by hundreds of Hz\citep{vdk05}.
In the Figure  \ref{fig:QPO}, show that Keck II spectroscopy of 
optical mHz quasi-periodic oscillations (QPOs) 
in the light curve of the X-ray pulsar binary Hercules X-1\citep{ohb+01}.

\begin{figure}
\includegraphics[bb=1.1cm 1.5cm 18cm 27cm,scale=0.40]{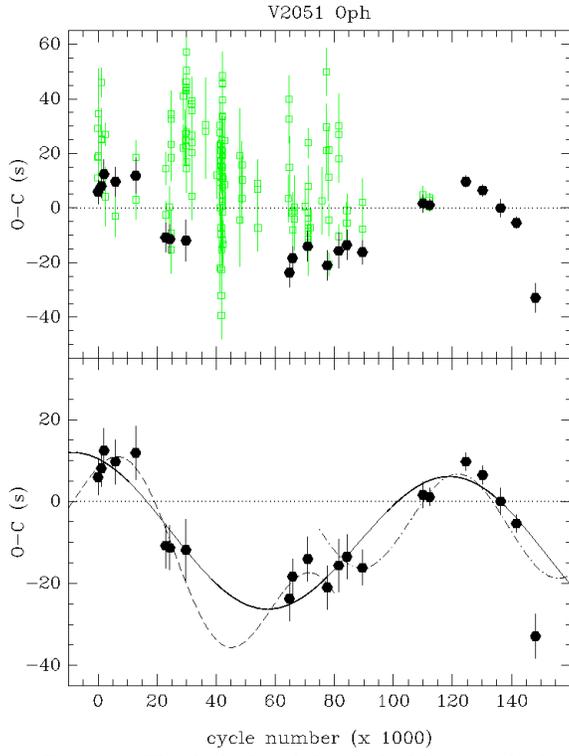}
\caption{The dashed and dot-dashed lines in the lower panel 
show the best-fit 11~yr cycle period sinusoidal ephemeris, 
respectively, for the data in the first and 
the second halves of the time interval. \citep[see][]{bbb+03}}
  \label{fig:cyclo}
\end{figure}

Year- to decade-long cyclic orbital period changes have been observed
in several classes of close binary systems, including Algol, W Ursae
Majoris, and RS Canum Venaticorum systems and the cataclysmic variables. 
The origin of these changes is unknown, but mass loss, apsidal motion, 
magnetic activity, and the presence of a third body have all been proposed.
In the Figure \ref{fig:cyclo}, show that cyclical period changes 
in the dwarf novae V2051 Oph.

The other two statistics periodic phenomenon also relate with stellar.
One is Hallinan, G. et al. detected periodic bursts of extremely bright, 
circularly polarized, coherent radio emission 
from the ultracool dwarf\citep{hbl+07}.
Another is Double Periodic Variables (DPVs) that are blue stars
characterized by a short periodicity (1-16 days) and
a long periodicity (50-600 days) in their light curves.
They were discovered in the Magellanic Clouds 
after a search for Be stars in the OGLE variable star catalog\citep{mpdg03}

\section{DISCUSSION AND CONCLUSIONS}
Signal detection techniques designed for cyclostationary signals 
take account of the periodicity or almost periodicity of 
the signal autocorrelation function. 
Single-cycle and multicycle detectors exploit one or multiple cycle frequencies,
respectively\citep{gnp06}.
Some search-efficient methods of detection of cyclostationary
signals\citep{yg96} and higher-order cyclostationary for weak-signal
detection\citep{sg92} be developed.
These will benefit of search transient source in the time variable universe, 
especially that have periodic in statistics.

Although no periodicities were detected in any of the sources using
standard Fourier or folding methods, 
for ten of the sources(RRATs) McLaughlin, M.A. et al. identify a periodicity
from the arrival times of the individual bursts
that used search techniques similar to those described in \citep{cm03}. 
In short, the 35-minute time series were dedispersed for a number of trial
values of DM.  The time series were smoothed by convolution with
boxcars of various widths to increase sensitivity to broadened pulses,
with a maximum boxcar width of 32~ms. Because the optimal sensitivity
is achieved when the smoothing window width equals the burst width,
our sensitivity is lower for burst durations greater than 32~ms.  Each
of these time series was then searched for any bursts above a
threshold of five standard deviations, computed by calculating a
running mean and root-mean-square deviation of the noisy time series.
All bursts detected above a 5-$\sigma$ threshold are plotted as
circles, with size proportional to the signal-to-noise ratio of the
detected burst. The abcissa shows arrival time while the ordinate
shows the DM.  Because of their finite width, intense bursts are
detected at multiple DMs and result in vertical broadening of the
features. Bursts which are strongest at zero DM and therefore likely
to be impulsive terrestrial interference are not shown. In general
these were easily identified by their detection in multiple beams of
the 13-beam receiver\citep{mll+06}.
More recently, Deneva, J.S. et al. use the above method 
and a friends-of-friends algorithm
perform the ongoing Arecibo Pulsar ALFA (PALFA) survey 
of the Galactic plane，then discover seven objects\citep{dcm+09} .

The discovery of RRATs increases the current Galactic population
estimates of radio pulsar by at least several times.
It seems also that there will be many candidates
for whom it will be impractical or impossible to follow up at
present with current observing facilities. These will require
followup with instruments like LOFAR, FAST or the SKA.
Keane, E.F. et al. note that these instruments will produce extraordinarily 
large volumes of data so that searching for transient RRAT-like sources 
will necessitate the development of automated algorithms which will 
use the steps as outlined above\citep{kle+09}.
So I think we should use the detection methods of cyclostationary
process integrate with DM search, 
that will be one automated and effective algorithms to seek RRATs.

If we do similar things in find the other astronomical cyclostationary sources
that include Intermittent Pulsars, QPOs et al., 
it will lead to discover more unusual astronomical phenomenon.

\acknowledgments
DJ thanks

\bibliographystyle{apj}
\bibliography{cyclo}

\begin{thebibliography}{27}
\expandafter\ifx\csname natexlab\endcsname\relax\def\natexlab#1{#1}\fi

\bibitem[{{Baptista} {et~al.}(2003){Baptista}, {Borges}, {Bond}, {Jablonski},
  {Steiner}, \& {Grauer}}]{bbb+03}
{Baptista}, R., {Borges}, B.~W., {Bond}, H.~E., {Jablonski}, F., {Steiner},
  J.~E., \& {Grauer}, A.~D. 2003, \mnras, 345, 889

\bibitem[{{Bretteil} \& {Weber}(2005)}]{bw05}
{Bretteil}, S., \& {Weber}, R. 2005, Radio Science, 40, 5

\bibitem[{{Camilo} {et~al.}(2006){Camilo}, {Ransom}, {Halpern}, {Reynolds},
  {Helfand}, {Zimmerman}, \& {Sarkissian}}]{crh+06}
{Camilo}, F., {Ransom}, S.~M., {Halpern}, J.~P., {Reynolds}, J., {Helfand},
  D.~J., {Zimmerman}, N., \& {Sarkissian}, J. 2006, \nat, 442, 892

\bibitem[{{Cordes} \& {McLaughlin}(2003)}]{cm03}
{Cordes}, J.~M., \& {McLaughlin}, M.~A. 2003, \apj, 596, 1142

\bibitem[{{Deneva} {et~al.}(2009){Deneva}, {Cordes}, {Mc Laughlin}, {Nice},
  {Lorimer}, {Crawford}, {Bhat}, {Camilo}, {Champion}, {Freire}, {Edel},
  {Kondratiev}, {Hessels}, {Jenet}, {Kasian}, {Kaspi}, {Kramer}, {Lazarus},
  {Ransom}, {Stairs}, {Stappers}, {van Leeuwen}, {Brazier}, {Venkataraman},
  {Zollweg}, \& {Bogdanov}}]{dcm+09}
{Deneva}, J.~S., {Cordes}, J.~M., {Mc Laughlin}, M.~A., {Nice}, D.~J.,
  {Lorimer}, D.~R., {Crawford}, F., {Bhat}, N.~D.~R., {Camilo}, F., {Champion},
  D.~J., {Freire}, P.~C.~C., {Edel}, S., {Kondratiev}, V.~I., {Hessels},
  J.~W.~T., {Jenet}, F.~A., {Kasian}, L., {Kaspi}, V.~M., {Kramer}, M.,
  {Lazarus}, P., {Ransom}, S.~M., {Stairs}, I.~H., {Stappers}, B.~W., {van
  Leeuwen}, J., {Brazier}, A., {Venkataraman}, A., {Zollweg}, J.~A., \&
  {Bogdanov}, S. 2009, \apj, 703, 2259

\bibitem[{Edlund {et~al.}(2005)Edlund, Tinto, Krolak, \& Nelemans}]{e05}
Edlund, J.~A., Tinto, M., Krolak, A., \& Nelemans, G. 2005, Phys. Rev., D71,
  122003

\bibitem[{Gardner(1978)}]{g78}
Gardner, W. 1978, IEEE Transactions on Information Theory, 24, 8

\bibitem[{Gardner(1991)}]{g91}
---. 1991, IEEE Transactions on Information Theory, 37, 216

\bibitem[{Gardner(1994)}]{g94}
---. 1994, {Cyclostationarity in communications and signal processing} (IEEE
  press New York)

\bibitem[{Gardner \& Franks(1975)}]{gf75}
Gardner, W., \& Franks, L. 1975, IEEE Transactions on Information Theory, 21, 4

\bibitem[{Gardner {et~al.}(2006)Gardner, Napolitano, \& Paura}]{gnp06}
Gardner, W., Napolitano, A., \& Paura, L. 2006, Signal processing, 86, 639

\bibitem[{{Hallinan} {et~al.}(2007){Hallinan}, {Bourke}, {Lane}, {Antonova},
  {Zavala}, {Brisken}, {Boyle}, {Vrba}, {Doyle}, \& {Golden}}]{hbl+07}
{Hallinan}, G., {Bourke}, S., {Lane}, C., {Antonova}, A., {Zavala}, R.~T.,
  {Brisken}, W.~F., {Boyle}, R.~P., {Vrba}, F.~J., {Doyle}, J.~G., \& {Golden},
  A. 2007, \apjl, 663, L25

\bibitem[{Hurd(1989)}]{h89}
Hurd, H. 1989, IEEE Transactions on Information Theory, 35, 350

\bibitem[{Hurd(1997)}]{h97}
---. 1997, Extract of Lectures Notes for the University of North Carolina

\bibitem[{{Keane} {et~al.}(2009){Keane}, {Ludovici}, {Eatough}, {Kramer},
  {Lyne}, {McLaughlin}, {Stappers}, \& {.}}]{kle+09}
{Keane}, E.~F., {Ludovici}, D.~A., {Eatough}, R.~P., {Kramer}, M., {Lyne},
  A.~G., {McLaughlin}, M.~A., {Stappers}, B.~W., \& {.} 2009, ArXiv e-prints

\bibitem[{{Kramer} {et~al.}(2006){Kramer}, {Lyne}, {O'Brien}, {Jordan}, \&
  {Lorimer}}]{klo+06}
{Kramer}, M., {Lyne}, A.~G., {O'Brien}, J.~T., {Jordan}, C.~A., \& {Lorimer},
  D.~R. 2006, Science, 312, 549

\bibitem[{{Leshem} {et~al.}(2000){Leshem}, {van der Veen}, \&
  {Boonstra}}]{lvb00}
{Leshem}, A., {van der Veen}, A., \& {Boonstra}, A. 2000, \apjs, 131, 355

\bibitem[{{Lyne}(2007)}]{lyne07}
{Lyne}, A. 2007, in Proceedings of ''Bursts, Pulses and Flickering: wide-field
  monitoring of the dynamic radio sky''. 12-15 June 2007, Kerastari, Tripolis,
  Greece., p.1

\bibitem[{{McLaughlin} {et~al.}(2006){McLaughlin}, {Lyne}, {Lorimer}, {Kramer},
  {Faulkner}, {Manchester}, {Cordes}, {Camilo}, {Possenti}, {Stairs}, {Hobbs},
  {D'Amico}, {Burgay}, \& {O'Brien}}]{mll+06}
{McLaughlin}, M.~A., {Lyne}, A.~G., {Lorimer}, D.~R., {Kramer}, M., {Faulkner},
  A.~J., {Manchester}, R.~N., {Cordes}, J.~M., {Camilo}, F., {Possenti}, A.,
  {Stairs}, I.~H., {Hobbs}, G., {D'Amico}, N., {Burgay}, M., \& {O'Brien},
  J.~T. 2006, \nat, 439, 817

\bibitem[{{Mennickent} {et~al.}(2003){Mennickent}, {Pietrzy{\'n}ski}, {Diaz},
  \& {Gieren}}]{mpdg03}
{Mennickent}, R.~E., {Pietrzy{\'n}ski}, G., {Diaz}, M., \& {Gieren}, W. 2003,
  \aap, 399, L47

\bibitem[{{O'Brien} {et~al.}(2001){O'Brien}, {Horne}, {Boroson}, {Still},
  {Gomer}, {Oke}, {Boyd}, \& {Vrtilek}}]{ohb+01}
{O'Brien}, K., {Horne}, K., {Boroson}, B., {Still}, M., {Gomer}, R., {Oke},
  J.~B., {Boyd}, P., \& {Vrtilek}, S.~D. 2001, \mnras, 326, 1067

\bibitem[{Serpedin {et~al.}(2005)Serpedin, Panduru, Sar{\i}, \&
  Giannakis}]{spsg05}
Serpedin, E., Panduru, F., Sar{\i}, I., \& Giannakis, G. 2005, Signal
  Processing, 85, 2233

\bibitem[{{Shabanova}(2009)}]{s09}
{Shabanova}, T.~V. 2009, \apj, 700, 1009

\bibitem[{Spooner \& Gardner(1992)}]{sg92}
Spooner, C., \& Gardner, W. 1992, in Proc. IEEE Sixth SP Workshop on
  Statistical Signal and Array Processing, 197--201

\bibitem[{{van der Klis}(2005)}]{vdk05}
{van der Klis}, M. 2005, Astronomische Nachrichten, 326, 798

\bibitem[{{van der Klis} {et~al.}(1985){van der Klis}, {Jansen}, {van
  Paradijs}, {Lewin}, {van den Heuvel}, {Trumper}, \& {Szatjno}}]{QPO85}
{van der Klis}, M., {Jansen}, F., {van Paradijs}, J., {Lewin}, W.~H.~G., {van
  den Heuvel}, E.~P.~J., {Trumper}, J.~E., \& {Szatjno}, M. 1985, \nat, 316,
  225

\bibitem[{Yeung \& Gardner(1996)}]{yg96}
Yeung, G., \& Gardner, W. 1996, IEEE Transactions on Signal Processing, 44,
  1214

\end{thebibliography}
\clearpage

\end{document}